\documentclass{webofc}
\usepackage[varg]{txfonts}   
\usepackage{color}

\begin{document}
\title{Exploring the freeze-out hypersurface of relativistic nuclear collisions with a rapidity-dependent thermal model}

\author{\firstname{Han} \lastname{Gao}\inst{1}\fnsep\thanks{\email{han.gao3@mail.mcgill.ca} (Speaker)} \and
        \firstname{Lipei} \lastname{Du}\inst{1}\fnsep\thanks{\email{lipei.du@mail.mcgill.ca}} \and
        \firstname{Sangyong} \lastname{Jeon}\inst{1}\fnsep\thanks{\email{sangyong.jeon@mcgill.ca}} \and
        \firstname{Charles} \lastname{Gale}\inst{1}\fnsep\thanks{\email{charles.gale@mcgill.ca}}
}

\institute{Department of Physics, McGill University, 3600 rue University, Montréal, QC, H3A 2T8, Canada
          }

\abstract{%
Considering applications to relativistic heavy-ion collisions, we develop a rapidity-dependent thermal model that includes thermal smearing effect and longitudinal boost. We calibrate the model with thermal yields obtained from a multistage hydrodynamic simulation. Through Bayesian analysis, we find that our model extracts freeze-out thermodynamics with better precision than rapidity-independent models. The potential of such a model to constrain longitudinal flow from data is also demonstrated.
}
\maketitle
\section{Introduction}
\label{intro}
Thermal models are powerful tools for extracting freeze-out (FO) thermodynamics from hadron yields \cite{Braun-Munzinger:2003pwq,Andronic:2008gu,Becattini:2003wp}. However, since boost invariance is violated at lower collisional energies \cite{Bjorken:1982qr}, it becomes crucial to incorporate both the thermal smearing effect and longitudinal flow in the thermal approach. In this study, such a model is developed.
A Bayesian analysis is performed by applying the model to hydrodynamic yields from simulations \cite{Du:2023gnv}, serving as a closure test for the model.
\section{Rapidity-dependent thermal model}
Assuming thermal equilibrium, the phase-space distribution of the hadron species $h = p,\pi^+,K^+$ for a FO cell at mid-rapidity ($y_s=0$) is given by its temperature ($T$), baryon chemical potential ($\mu$) and volume ($V$)
\begin{equation} \label{eq:dnd3p}
    \left.\frac{dN^h}{d^3\vec p}\right|_{y_s=0} = \frac{g_h V}{(2\pi)^3} f_h(E_{\vec p}).
\end{equation}
Here, $f_h = \frac{1}{e^{(E_p - \mu_h)/T} \pm 1}$ is the Fermi-Dirac or Bose-Einstein distribution of the hadron, and $g_h$ is the spin degeneracy. It is worth noticing that Eq. (\ref{eq:dnd3p}) matches with the Cooper-Frye (CF) prescription without non-equilibrium correction \cite{Du:2023gnv}. By integrating out transverse momentum, the rapidity distribution of hadrons for this cell is given by
\begin{equation}
\label{eq:pik_yield}
    \left.\frac{dN^h}{dy}\right|_{y_s=0} \equiv   K^h(y;T,\mu,V) =\frac{g_h VT^3}{(2\pi)^2}\sum_{n=1}^\infty\frac{1}{n^3} \left( \frac{2}{\cosh^2 y} + \frac{nm_h}{T}\frac{2}{\cosh y} + \frac{n^2m^2_h}{T^2}\right)e^{-\frac{nm_h\cosh y}{T}}.
\end{equation}
Eq. \ref{eq:pik_yield} suggests that a FO cell with a given rapidity $y_s$ non-locally contributes hadron yields to the entire rapidity range, which represents the thermal smearing effect \cite{Schnedermann:1993ws,Begun:2018efg}.
\par 
The broken boost invariance results in the rapidity of a freeze-out cell, $y_s$, being larger than its space-time rapidity, $\eta_s$. For a given flow velocity $\tau u^\eta$, they are related by\begin{equation}\label{eq:yetas}
    y_s(\eta_s) = \frac{1}{2}\ln \frac{( \sqrt{ 1 + (\tau u^\eta)^2} + \tau u^\eta)(1+\tanh \eta_s)}{( \sqrt{ 1 + (\tau u^\eta)^2} - \tau u^\eta)(1-\tanh \eta_s)}.
\end{equation}
In this study, the velocity is parametrized as $\tau u^\eta = \alpha \eta_s^3$.
\par 
Given the functional dependence of freeze-out thermodynamics along the space-time rapidity, namely $T(\eta_s), V(\eta_s)$ and $\mu(\eta_s)$, the thermal yields of a hadron species $h$ is obtained by an integral over all freeze-out cells
\begin{equation}\label{eq:dndy}
    \frac{dN^h}{dy} = \int_{-\eta_{\rm max}}^{\eta_{\rm max}} d\eta_s K^h(y-y_s(\eta_s);T(\eta_s),V(\eta_s),\mu(\eta_s)).
\end{equation}
Here, $\eta_{\rm max}$ refers to the maximal range that the FO hypersurface extends in the $\eta_s$ space. Eqs. (\ref{eq:pik_yield}-\ref{eq:dndy}) together constitute the rapidity-dependent thermal model.

\section{Testing the thermal model: Bayesian analysis of the freeze-out thermodynamics}
The decay of resonant states and heavy particles can contribute further to the measured final yields; this is known as the feed-down effect. Since this effect is not included in Eq. (\ref{eq:dndy}), we utilize of a multistage hydrodynamic simulation presented in Ref. \cite{Du:2023gnv}, where both the final and thermal (CF) yields are available.
\par 
The quantities characterizing the FO thermodynamics are parametrized as
\begin{align}
      T(\eta_s) &= T_0 + T_2 \eta_s^2,\label{eqs:paramT} \\
      V(\eta_s) &= V_0 + V_2\eta_s^2,\label{eqs:paramV} \\
      z(\eta_s) &= \exp\left[\frac{\mu(\eta_s)}{T(\eta_s)}\right] - \exp\left[-\frac{\mu(\eta_s)}{T(\eta_s)}\right] = z_0 + z_2 \eta_s^2.\label{eqs:paramZ}
\end{align}
\par
To find how the data constrain the FO thermodynamics and longitudinal flow, a Bayesian analysis is conducted for the parameter set $\theta = (T_0,T_2,V_0,V_2,z_0,z_2,\alpha,\eta_{\rm max})$ with respect to the thermal hadron yields. Denoting the model-predicted yields at rapidity $y$ given the thermal model parameter set $\theta$ as $\frac{dN^h}{dy}(y;\theta)$, we use a Gaussian likelihood function given by
\begin{equation}
    L(\theta) = \prod_{h=p-\bar p,\pi^+,K^+}\prod_{y_i} \frac{1}{\sqrt{2\pi}\epsilon_i^h} \exp\left[ -\left(\frac{dN^h}{dy}(y_i;\theta) - \frac{dN^h_{\rm CF}}{dy}(y_i)\right)^2/(2\epsilon_i^2)\right],
\end{equation}
where $\frac{dN^h_{\rm CF}}{dy}(y_i)$ denotes the thermal yields from the hydrodynamic simulation for hadron species $h$, and $\epsilon_i^h = \frac{dN^h_{\rm CF}}{dy}(y_i)\times 5\%$ represents the estimated error for the thermal yields.
Note that $y_i\in [0,4]$ are the sampled rapidity points, chosen to be dense around mid-rapidity ($y\lesssim 1$) but sparse for large rapidity, mimicking experimental measurements \cite{BRAHMS:2009wlg,STAR:2017sal}. \par
The prior distribution of the parameter set $p(\theta)$ listed in Table \ref{tab:prior} is a conservative estimation based on our previous knowledge from both experiments and hydrodynamic simulations.
\begin{table}[h]
\centering
\caption{A list of all used prior distributions for 0-5\% Au+Au collision at $\sqrt s = 19.6{\rm \ GeV}$. For non-negative parameters, the inverse gamma distribution is employed, covering an interval with a $99\%$ probability \cite{Heffernan:2021klw}. Parameters that can take both positive and negative values are assigned a Gaussian prior distribution, where the interval is defined by $\mu \pm 3\sigma$ values. The parameters are referenced in Eqs. \ref{eqs:paramT}, \ref{eqs:paramV}, \ref{eqs:paramZ}, and \ref{eq:dndy}.}
\label{tab:prior}       
\begin{tabular}{ll|ll}
\hline
Parameter & Interval & Parameter & Interval \\\hline
$T_0$ & [0.12,0.17] GeV & $T_2$ &[-0.05,0.05] GeV\\
$V_0$ & [600,6000] fm$^3$ &$V_2$ & [-1500,1500] fm$^3$\\
$z_0$ & [0.05,2] & $z_2$ & [-2,2] \\
$\alpha$ & [-0.1,0.1] & $\eta_{\rm max}$ & [1,4]
\\\hline
\end{tabular}
\end{table}
Given the likelihood and the prior distribution, the posterior distribution of the parameters is derived by the Bayesian theorem $p(\theta| dN_{\rm CF}/{dy}) \propto L(\theta)p(\theta)$. To sample $\theta$ from this distribution, a Monte Carlo Markov Chain sampling is employed.

\section{Results and discussions}
Bayesian analysis are performed for 0-5\% Au+Au collision at beam energy $\sqrt s = 7.7,19.6,62.4$ and $200{\rm \ GeV}$ in the corresponding simulation. Here, we present the results for  $\sqrt s = 19.6{\rm \ GeV}$.
\begin{figure}[h]
\centering
\includegraphics[width=6cm,clip]{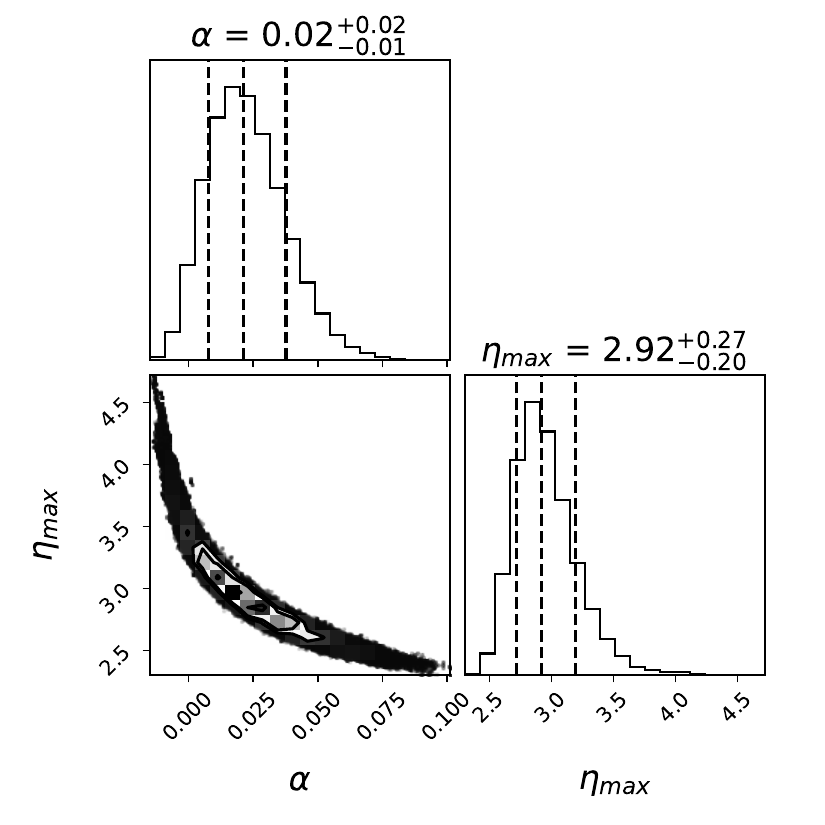}
\includegraphics[width=6cm,clip]{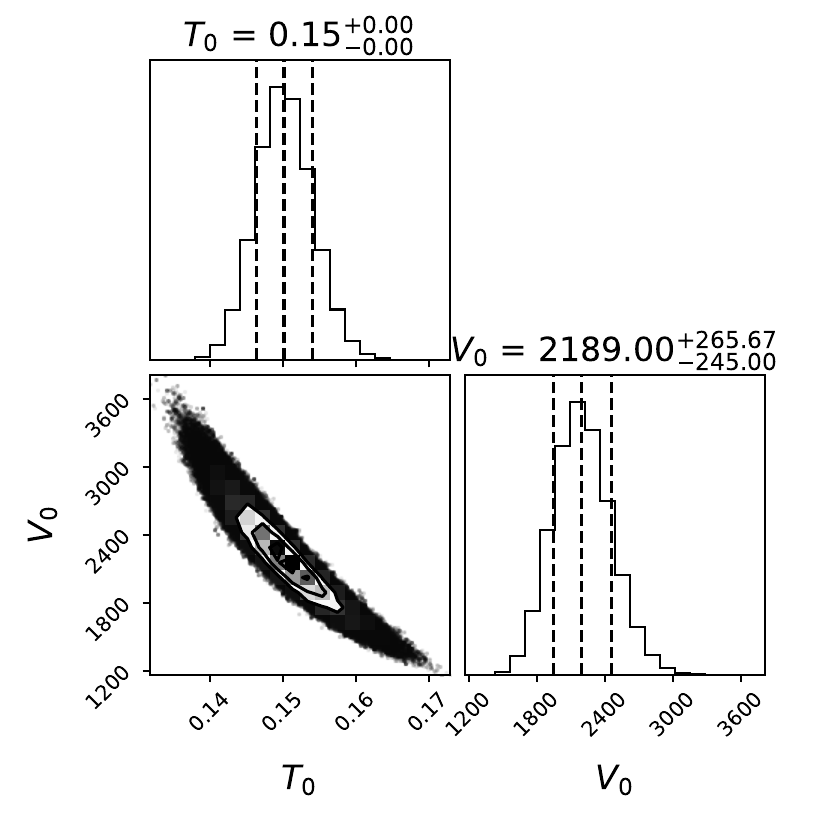}
\caption{Posterior distribution of parameters $(\alpha,\eta_{\rm max},T_0,V_0)$ listed in table \ref{tab:prior}, for 0-5\% Au+Au collision at $\sqrt s = 19.6{\rm \ GeV}$. A strong correlation is noticed between $\alpha$ and $\eta_{\rm max}$, as well as between $T_0$ and $V_0$. The dashed lines indicate $16\%,50\%$ and $84\%$ percentiles for the parameter distribution.}
\label{corr}       
\end{figure}
\par
\emph{Longitudinal system size and flow strength} --
In the left panel of Fig. \ref{corr}, we observe a significant correlation between the strength of the longitudinal flow $\alpha$ and the longitudinal system size $\eta_{\rm max}$. 
However, the Bayesian analysis still suggests a positive $\alpha$, which is indeed found in the corresponding hydrodynamic simulation. This suggests that a direct constrain on the longitudinal flow is indeed accessible from thermal models.
\par
\emph{Mid-rapidity temperature and volume} -- 
Another significant correlation shown in the right panel of Fig.\ref{corr} is the volume $V_0$ and temperature $T_0$ at mid-rapidity. This reflects the conservation of the total system entropy $S\sim VT^3$. Nevertheless, the model still gives small uncertainties in quantifying $T_0$ and $V_0$. 
\begin{figure}[h]
\centering
\sidecaption
\includegraphics[width=7cm,clip]{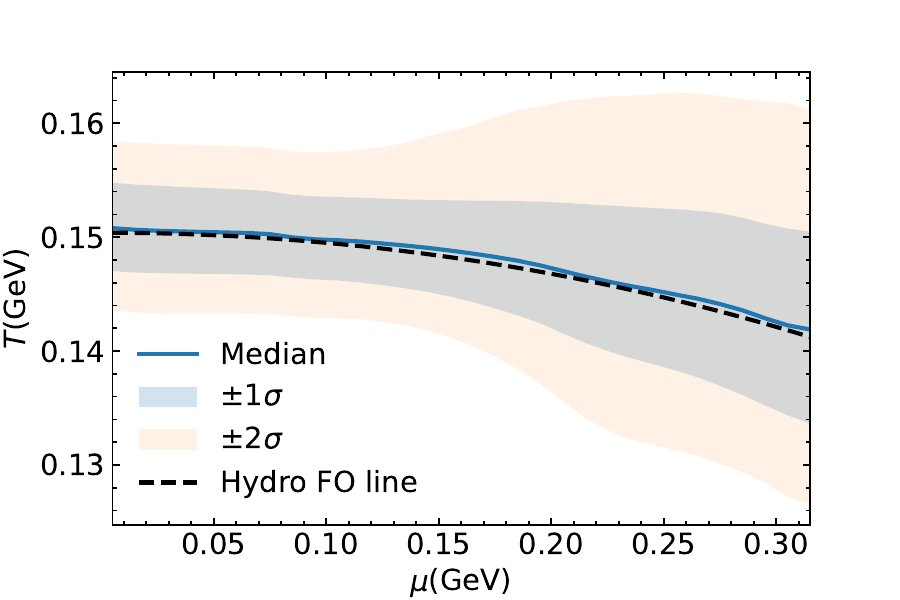}
\caption{Freeze-out distribution drawn from the rapidity-dependent thermal model. The dashed black line represents the freeze-out condition used in the corresponding hydrodynamic simulation. As a good match is found between the median and the hydrodynamic FO condition, it is suggested that the model gives the FO thermodynamics with good precision.}
\label{fig:pd}       
\end{figure}
\par 
\emph{Freeze-out thermodynamics and phase diagram} -- 
In Fig. \ref{fig:pd}, we show the distribution of $(T,\mu)$ for all sampled FO cells. As the model is applied to yields from a hydrodynamic simulation where a FO condition is well-defined, comparing the samples with the hydrodynamic FO condition serves as a closure test for the model. A good match is indeed observed in Fig. \ref{fig:pd}. Therefore, our rapidity-dependent thermal model demonstrates better accuracy in extracting FO thermodynamics than rapidity-independent models \cite{Du:2023gnv}.

\section{Conclusion}
As our thermal model includes the thermal smearing and longitudinal flow, its ability to accurately and consistently explore the FO thermodynamics is well demonstrated in the presented test case. As $\sqrt s$ decreases, the FO hypersurface evidently becomes less isothermal \cite{Du:2023gnv}, the need for considering these two effects in the thermal model then becomes imperative. It will be interesting to incorporate the feed-down effect and apply our model directly to experimental measurements.

\vspace{0.5cm}
\noindent\textbf{Acknowledgements:} This work was supported in part by the Natural Sciences and Engineering Research Council of Canada. Computations were made on the B\'eluga computer managed by Calcul Qu\'ebec and by the Digital Research Alliance of Canada.

\end{document}